\documentclass[12pt]{article}
\usepackage{graphicx}
\pdfoutput=1
\usepackage{graphicx}
\usepackage{dcolumn}
\usepackage{comment}
\usepackage{physics}
\usepackage{amsmath}
\usepackage{multirow}
\usepackage{amsmath}
\usepackage{diagbox}
\usepackage{tkz-graph}
\tikzset{
  LabelStyle/.style = { rectangle, rounded corners,
                        minimum width = 1em, fill = white!50,
                        text = black, font = \bfseries },
  VertexStyle/.append style = { inner sep=2pt,
                                font = \bfseries},
  EdgeStyle/.append style = {->,} }
\pdfoutput=1
\usepackage[utf8]{inputenc}
\usepackage[english]{babel}
\usepackage[T1]{fontenc}
\usepackage{amsmath}
\usepackage{hyperref}
\usepackage{physics}
\usepackage{multirow}
\usepackage{amssymb}
\usepackage{colortbl}   
\usepackage{diagbox}
\usepackage{tkz-graph}
\tikzset{
  LabelStyle/.style = { rectangle, rounded corners,
                        minimum width = 1em, fill = white!50,
                        text = black, font = \bfseries },
  VertexStyle/.append style = { inner sep=2pt,
                                font = \bfseries},
  EdgeStyle/.append style = {->,} }
\usepackage{bm}

\usepackage{tikz}
\usepackage{lipsum}

\usepackage{authblk}
\begin{document}
\font\myfont=cmr12 at 24pt

\title{{\myfont Quantum strategies for simple 2-player XOR games}}
\author[$\;$]{Ricardo Faleiro}
\affil[$\;$]{Instituto de Telecomunicaç\~oes and Departamento de Matem\'atica\\  Instituto Superior T\'ecnico, Avenida Rovisco Pais 1049-001, Lisboa, Portugal}
\date{}                     
\renewcommand\Affilfont{\itshape\small}

\maketitle

\begin{abstract}
 The non-local game scenario provides a powerful framework to study the limitations of classical and quantum correlations, by studying the upper bounds of the winning probabilities those correlations offer in cooperation games where communication between players is prohibited. Building upon results presented in the seminal work of Cleve et al. \cite{Cleve}, a straightforward construction to compute the Tsirelson bounds for simple 2-player XOR games is presented. The construction is applied explicitly to some examples, including the \textit{Entanglement Assisted Orientation in Space} (EAOS) game of Brukner et al. \cite{EAOS}, proving for the first time that their proposed quantum strategy is in fact the optimal, as it reaches the Tsirelson bound.
\end{abstract}

\section{Introduction}
Non-signaling games are cooperation multiplayer games where the players do not know all the information they could know in order to play the game in an ideal manner -- they only know explicitly the information that was given to them by a neutral party, appropriately entitled as the \textit{Referee}. This is usually imposed by a constraint called the \textit{No Signaling Condition}, where communication either classical or quantum is not allowed between the players (physically one could think that the players are \textit{spacelike} separated from one another). This type of game is called \textit{non-local} when players using strategies that exploit the non-locality of quantum mechanics, i.e \textit{quantum strategies}, can reach higher probabilities to win than players restricted to using \textit{classical strategies}. A short and concise overview on non-local games can be found in \cite{Buhrman}. Such games always evolve according to the following stages\footnote{These strictly speaking are the stages of just one \textit{round} of a game and some games could have more than one round -- since this work deals exclusively with one round games this description fully characterizes the evolution of such games.},
\begin{itemize}
    \item The Referee sends to each player a specific input, usually referred to as a \textit{question} ($q$);
     \item Each player only receives its own \textit{question} and since they can't communicate with one another they are ignorant of the others'. Then each player will produce an output i.e an \textit{answer} ($a$) based on a  previously agreed common strategy and send them to the Referee;
     \item The Referee will check  the players' \textit{answers} against the \textit{questions} and see if they are ``correct" i.e if they respect the winning condition  specified in the rules of the game;
\end{itemize}

Also, depending on whether the \textit{questions} and/or \textit{answers}  used in the game are classical or quantum information, we say the game is a \textit{ classical non-local game} or \textit{quantum non-local game}, respectively. This work deals with {classical non-local games}, which means that the \textit{questions} and \textit{answers} are classical information -- this does not imply, however, that the strategies should be exclusively classical. {Quantum strategies} which are strategies that exploit entangled quantum states can be used in the context of classical non-local games because the states are never explicitly communicated, they are just measured.

\subsection{Notation and definitions}

Since the \textit{questions} and \textit{answers} are classical, they are represented mathematically as elements of sets. Keeping the standard terminology of upper case for sets and lower case for elements of the set we say that $Q_i$ and $A_i $  are, respectively, the set of all \textit{questions} and \textit{answers} the $i$th player can receive. Similarly, $q_i$ and $a_i$ are the \textit{question} and \textit{answer} the $i$th player actually received in a run of the game. If we are dealing with a $n$-player ($n\geq$2)  non-local game, then\begin{equation*} 
\begin{split}
  Q=Q_1  \times \ldots \times Q_i \times \ldots\times Q_n; \;\;
   A=A_1  \times \ldots \times A_i \times \ldots\times A_n;
\end{split}
\end{equation*}are, respectively, the set of all the possible  \textit{questions} the players can receive, and \textit{answers} they can give. They are mathematically the Cartesian products of each players individual set of \textit{questions} and \textit{answers}. Accordingly we have that a general element of both of the previous sets, $(q_1   \ldots  q_i  \ldots q_n) \in Q$ and $(a_1   \ldots  a_i  \ldots a_n) \in A$, are the combination of \textit{questions} the  Referee gave and the \textit{answers} he received in return from the $n$ players.

Any given $n$-player ($n\geq$2) non-local game is completely defined by,

\begin{itemize}
    \item A probability distribution, which specifies how likely the Referee is to ask any given combination of questions to the players, \begin{center}
  $p(q_1, \ldots ,q_i,\ldots,q_n)$;
\end{center}{}
\item The \textit{predicate}, which is a Boolean function that outputs either $0$ or $1$ depending on its input. The input is some  ending game configuration i.e a pair of the form \{$(q_1,\ldots,q_i,\ldots,q_n),(a_1,\ldots,a_i,\ldots,a_n)$\}, and it evaluates to $1$ if the configuration wins the game and to $0$ if it loses. The predicate is usually written like,

\begin{center}
    $V(a_1,\ldots,a_i,\ldots,a_n|q_1,\ldots,q_i,\ldots,q_n) .$
\end{center}{}
in the spirit of a conditional probability, to illustrate that the validity of the \textit{answers}  is conditioned on the \textit{questions}.
\end{itemize}

Now let us adopt the following short-hand notation, $Q^x = (q_1   \ldots  q_i  \ldots q_n)$ if $(q_1   \ldots  q_i  \ldots q_n)$ is the $x$th element of  set $Q$ according to some specific order and $A^y=(a_1   \ldots  a_i  \ldots a_n)$ if $(a_1   \ldots  a_i  \ldots a_n)$ is the $y$th element of set $A$ according to the same type of order\footnote{Note that $Q^{x}$ does not mean the set of all possible questions for the $x$th player, like $Q_i$ meant for the $i$th player. It means the $x$th element of $Q$, whatever it might be according to some arbitrary order. Likewise $A^{y}$ means the $y$th element of $A$.}. There is nothing fundamental in this, it's just for purposes of increased readability in the expressions. Now we have that some non-local game $G$ is given by,

\begin{center}
  $p(Q^x)$ and $V(A^y|Q^x) $,
\end{center}{}

and to show explicitly that a non-local game $G$ is defined by just these two things, it is usually written as $G(V,p)$.
 According to this notation the predicate is symbolically given by,
\begin{equation*}
    V(A^y|Q^x) = \left\{
\begin{array}{ll}
      1, \hspace{.1cm}\textup{if }\hspace{.1cm} \{Q^x, A^y\}\; \textup{is a winning configuration }\\
      0, \hspace{.1cm}\textup{otherwise}\\
\end{array} 
\right.
\end{equation*}

\subsection{ \textit{Strategies} for non-local games}
A \textit{strategy} $\mathcal{S}$ specifies the probability function $p(A^y|Q^x)$, for every combination of $x$ and $y$. That is, the probability that the players will give a specific combination of \textit{answers} upon being asked a specific combination of \textit{questions}. It is not difficult to see that the probability to win some game $G=(V,p)$ with \textit{strategy} $\mathcal{S}$, is given by the expectation value of the probabilities to reach all possible configurations $\{Q^x, A^y\}$ allowed by $\mathcal{S}$ and evaluated by the predicate $V$. We write, 
\begin{equation}
 \textup{W}_{\mathcal{S}}(G)  =  \sum_{x,y} p(Q^x) \; p(A^y|Q^x)\; V(A^y|Q^x),
    \label{ProbS}
\end{equation} where $\textup{W}_{\mathcal{S}}(G)$ is to be read as ``\textit{the probability to win game $G$ by using \textit{strategy} $\mathcal{S}$}". A good \textit{strategy} $\mathcal{S}$ is one which tries to maximize (\ref{ProbS}). Obviously finding the best \textit{strategy} would be trivial if communication was allowed, but in the context of non-local games, since that isn't the case, players only know their own \textit{questions} and the probability distribution $p(Q^x)$, so they are aware of how likely it is for the Referee to ask a specific  combination of \textit{questions}, but they don't know any other \textit{question} aside their own when playing the game -- this makes for a harder case.

\subsubsection{\textit{Classical Strategies} and Bell inequalities} A \textit{classical strategy} $\mathcal{C}$ could be either \textit{deterministic} or \textit{non-deterministic}. In a deterministic strategy the \textit{answers} are always given by a function of the form, $$A^y=F(Q^x),\;\;F(Q^x)\equiv f_1(q_1) \ldots f_n (q_n).$$ A non-deterministic strategy is just a probabilistic distribution over deterministic ones, so we have $$A^y= F_i(Q^x), \;\;\textup{with probability} \;\;p_i,$$  where $i$ is the index spanning the set of the deterministic strategies under consideration. It is easy to see that a deterministic strategy is the special case of the non-deterministic one where $p_i=1$ for some $i$. Perhaps not so immediate, but also true, is that you can find a  deterministic strategy that behaves at least as good as the best non-deterministic one. This is because since a non-deterministic strategy is the probabilistic distribution over a set of deterministic strategies, we can just pick the best one out of that set \footnote{This is equivalent to saying that the average over a set of positive numbers is never greater than  the highest number of the set. }. Then, we shall assume without loss of generality that the strategy $\mathcal{C}$ is deterministic, and as such we will substitute $A^y=F(Q^x)$ in (\ref{ProbS}) to get,

\begin{equation}
\begin{split}
\textup{ W}_  \mathcal{C}(G)
\end{split}
     =\sum_{x} p(Q^x)\; p(F(Q^x)|Q^x)\; V(F(Q^x)|Q^x).
    \label{PWinS}
\end{equation}

Since on input $Q^x$ the output will always be the one defined by $F(Q^x)$, it becomes evident that $p(F(Q^x)|Q^x)=1$, so,

\begin{equation}
\begin{split}
\textup{ W}_  \mathcal{C}(G)
\end{split}
     =\sum_{x} p(Q^x)V(F(Q^x)|Q^x).
    \label{PWinS2}
\end{equation}

The best { classical strategy} $\mathcal{C}^*$ is the one that maximizes the winning probability in (\ref{PWinS2}), then

\begin{equation}
\begin{split}
\textup{ W}_\mathcal{C^*}(G)
\end{split}
     \equiv \underset{F}{\textup{Max }}\;\sum_{x} p(Q^x) V(F(Q^x)|Q^x),
    \label{PWinS3}
\end{equation}

is the highest possible probability to win a given non-local game $G$, by means of a classical strategy, and is called the \textit{classical value}  of the game. In the literature it is usually depicted as $\omega_c(G)$. The following inequality holds true for any non-local game $G$,
\begin{equation}
\begin{split}
\textup{ W}_  \mathcal{C}(G)
\end{split}
     \;\;\leq \;\omega_c(G).
    \label{Bell}
\end{equation}

This is called a Bell inequality.

\subsubsection{\textit{Quantum Strategies} and Tsirelson inequalities} A \textit{quantum strategy} $\mathcal{Q}$, in the context of non-local games, is usually assumed to be a strategy that adds an extra resource which players can use, namely, {quantum entanglement}. $\mathcal{Q}$ is then defined by a finite dimensional entangled state $\ket{\psi}\in \mathcal{H} =\mathcal{H}_1 \otimes \ldots \otimes \mathcal{H}_n$ shared over all n players,  and a POVM for each $k$ player, $$
   \hat{\Pi}_k \equiv \big\{ \forall_{q_k \in Q_k}: \hat{\Pi}_{(q_k,a_k)}\big\}
$$


This means that player $k$ has the POVM defined for every possible input $q_k$ in such a way, that the measurement outcome of this POVM on the state $\ket{\psi}$ will give him answer $a_k$ with some probability. Then, the collection of questions $Q^x = q_1 \ldots q_n$ will define \begin{equation}
\label{probq}
    \hat{\Pi}_1 \otimes \ldots \otimes \hat{\Pi}_n \equiv \hat{\Pi}^x,
\end{equation}
in such a way that  the measurement outcome of $\hat{\Pi}^x$ will yield $A^y= a_1\ldots a_n$ with probability $\bra{\psi}(\hat{\Pi}^x)^{\dagger}(\hat{\Pi}^x)\ket{\psi}$. Which is to say that $$p(A^y|Q^x)=\bra{\psi}(\hat{\Pi}^x)^{\dagger}(\hat{\Pi}^x)\ket{\psi}.$$ Then for some {quantum strategy} $\mathcal{Q}$, (\ref{ProbS}) becomes

\begin{equation}
    W_{\mathcal{Q}}(G) =  \sum_{x} p(Q^x)\; \bra{\psi}(\hat{\Pi}^x)^{\dagger}(\hat{\Pi}^x)\ket{\psi} \; V(A^y|Q^x).
    \label{ProbQ}
\end{equation}

Similarly to the classical case, the best quantum strategy $Q^{*}$ is the one that maximizes (\ref{ProbQ}). Then we have that 

\begin{equation}
    \textup{ W}_  \mathcal{Q^*}(G)  \equiv \underset{\hat{\Pi}^x, \ket{\psi}}{{\textup{Max }}}\sum_{x} p(Q^x)\; \bra{\psi}(\hat{\Pi}^x)^{\dagger}\hat{\Pi}^x\ket{\psi} \; V(A^y|Q^x),
    \label{ProbQ*}
\end{equation}

is the highest possible probability to win a given non-local game $G$, by means of a quantum strategy, and is called the \textit{quantum value} of the game. It is usually depicted as $\omega_q(G)$. The following inequality holds true for any non-local game $G$,\begin{equation}
\begin{split}
\textup{ W}_  \mathcal{Q}(G)
\end{split}
     \;\;\leq \;\omega_q(G).
    \label{Tsirelson}
\end{equation}

This is called a Tsirelson inequality. 

\subsubsection{Non-local and pseudo-telepathy games} In the context of non-local games, the Bell and Tsirelson inequalities define the upper bounds on the winning probabilities, achieved by classical and quantum strategies. The distinct characteristic of a non-local game $G$ is then mathematically represented as \begin{equation}
\omega_c(G) < \;\omega_q(G),
    \label{correlations}
\end{equation}

which is the mathematical representation of what was previously stated -- a non-local game is a non-signaling game where the best quantum strategy always achieves a higher winning probability than the best classical strategy. This is not to say that quantum strategies are generally the optimal strategies in non-local games, a different type of strategies using another class of resources appropriately entitled \textit{non-local boxes}, or \textit{PR boxes} were engineered to be the best possible strategy for these types of games \cite{Broadbent}.

Interestingly, there is a special type of non-local game where the Tsirelson inequality is bounded by 1, which is to say that the best quantum strategy is the overall optimal strategy, since using the best quantum strategy will win the game with certainty i.e \begin{equation}
\omega_c(G) < 1 \wedge \;\omega_q(G) =1.
    \label{Telepathy}
\end{equation}

This type of non-local game is called a \textit{pseudo telepathy} game \cite{Brassard}. The name was chosen to illustrate the fact that if the Referee was ignorant to the possibility of quantum strategies, that the only possible explanation for Alice and Bob being able to always win the game would be to assume that they would have to be connected by some sort of illicit telepathic channel, that worked around the \textit{No Signaling Condition}. Some examples of this type of game are the Magic Square Game \cite{Aravind}, the Kochen-Specker Game \cite{Cleve} and also the Simple Game \cite{Brassard}.

\section{2-player non-local games}

From this section onward we shall be dealing exclusively with classical 2-player non-local games -- the players are the archetypal Alice and Bob, and we adopt the conventional nomenclature where Alice is asked question $s \in S$ and gives answer $a\in A$, and Bob is asked question  $t \in T$ and gives answer $b\in B$. The following table relates $n$-player to the 2-player game nomenclature,

\begin{center}
\begin{tabular}{ c | c }
  \hline \hline			
  $n$-player game & 2-player game \\
  \hline \hline
   $Q^x=(q_1, \ldots, q_n)$ & $Q^x=(s,t)$ \\  \\
   
    $A^y=(a_1, \ldots, a_n)$ & $A^y=(a,b)$ \\ \\
   
   $p(Q^x)= p(q_1, \ldots, q_n)$ & $p(Q^x)=p(s,t)$ \\  \\
   
   $V(A^y|Q^x)$ & $V(ab|st)$  \\ \\
   
    $\hat{\Pi}^x =  \hat{\Pi}_1 \otimes \ldots \otimes \hat{\Pi}_n  $ & $\hat{\Pi}^x = \hat{\Pi}_A\otimes \hat{\Pi}_B$ \\
    
  \hline  
\end{tabular}
\end{center}{}

Figure 1 is an illustration of how the 2-player game proceeds. The game goes as follows -- the Referee selects according to a probability distribution $p(s,t)$, question $s\in S$ to send Alice and question $t\in T$ to send Bob. Alice and Bob at that point know $p(s,t)$ and their own  respective questions, and choose their answers based on some preferred strategy, which is one that maximizes the winning probability\footnote{We make the implicit assumption that Alice and Bob always try to win the game with the highest possible probability.}. If they are using a classical strategy, they have to pick a function $F(s,t)$ that maximizes (\ref{PWinS2}), on the other hand, if they are using a quantum strategy, they have to choose a state $\ket{\psi}$ and two POVM's $\{\hat{\Pi}_A,\,\hat{\Pi}_B\}$ that maximize (\ref{ProbQ}).

\begin{center}
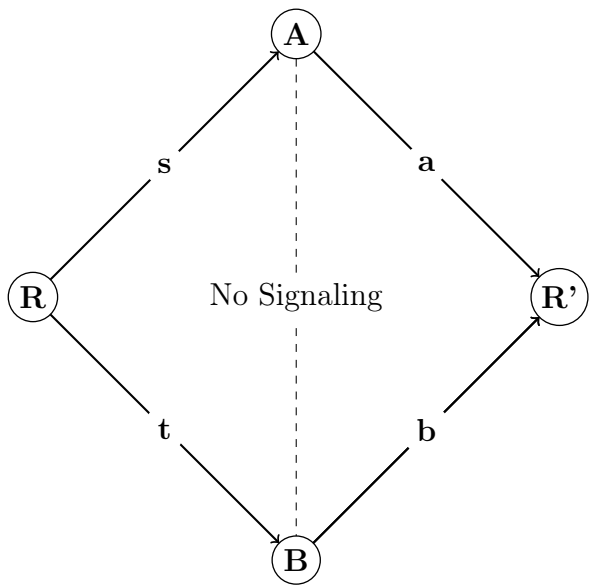
\begin{figure}[h!]
\begin{tikzpicture}
\hspace{3.25cm}
  \SetGraphUnit{3.5}
  {\tikzset{VertexStyle/.append style = {shape=coordinate}}
  \Vertex{No Signaling}}
  \WE(No Signaling){R}
  \EA(No Signaling){R'}
  \NO(No Signaling){A}
  \SO(No Signaling){B}
  \tikzset{EdgeStyle/.append style = {}}
  \Edge[label = t](R)(B)
  \tikzset{EdgeStyle/.append style = {}}
  \Edge[label = s](R)(A)
  \tikzset{EdgeStyle/.append style = {}}
  \Edge[label = a](A)(R')
  \tikzset{EdgeStyle/.append style = {}}
  \Edge[label = b](B)(R')
  \tikzset{EdgeStyle/.append style = {}}
  \Edge[label = b](B)(R')
  \draw[dashed] (A) -- (B) node[midway,fill=white] {No Signaling };
\end{tikzpicture}
  \caption{Single round non-local game between Alice (A) and Bob (B) mediated by the Referee; $s$ and $t$ are the questions  asked by the Referee at the start (R), to Alice and Bob respectively; $a$ and $b$ are the answers that Alice and Bob give back to the Referee at a later time (R'); The diagram could be interpreted either as an abstract graph or a spacetime diagram; }
    \label{NLG}
\end{figure}
\end{center}

\vspace{-1cm}

The graph in Fig.1 is the DAG (Directed Acyclic Graph) that represents how the game unfolds throughout time. The DAG could just be thought as an abstract graph showing an interactive picture of the game, or could actually be interpreted as being embedded in a Minkowski spacetime, thus being promoted to the spacetime diagram of the game (where the time arrow points from left to right) in $1\oplus1$ dimensions i.e $1$ of space plus $1$ of time. If we think about the diagram in the latter terms, the edges become \textit{wordlines} and the nodes become \textit{spacetime events} -- in that scenario the \textit{No Signaling Condition} would not need to be stated explicitly, as it comes naturally  from the geometry of the spacetime diagram, since  there is no way for Alice and Bob to communicate without the message passing first trough the Referee.

\subsection{XOR games}
XOR  games are a particular interesting type of non-local games, as they represent one of the few classes of games for which general upper bounds are known -- first introduced in \cite{Cleve},
XOR games are a subset of yet a larger set of non-local games entitled \textit{binary} games, in which the players only answer with bits to the Referee, even though the questions themselves need not be bits. A XOR game still restricts the set of binary games by specifying a special type of predicate -- we then say that a given 2 player non-local game $G$ is said to be a XOR  game if the {answers} $a$ and $b$ are \textit{bits} (i.e $G$ is a binary game) and the predicate of the game is given by, 

\begin{equation}
 V(ab| st)_{\textup{XOR}} =\left\{
\begin{array}{ll}
      1, \hspace{.1cm}\textup{if }\hspace{.1cm}f(s,t)=a\oplus b\\
      0, \hspace{.1cm}\textup{otherwise}\\
\end{array} .
\right.    
\end{equation}{}

This means that the winning condition of a XOR game does not depend explicitly on the outputs of the players but only on their parity, i.e whether the bits are the same or not. This is mathematically represented by the exclusive OR logical operation (which is just addition modulo 2) shortened as XOR. One example of a 2-player XOR game is the famous CHSH$(V,p)$ game, in which the questions $s,t$ are also bits. The game is defined by,

\begin{equation}
\label{predicateCHSH}
V(ab| st)_{\textup{CHSH}} =\left\{
\begin{array}{ll}
      1, \hspace{.1cm}\textup{if }\hspace{.1cm}s\cdot t=a\oplus b\\
      0, \hspace{.1cm}\textup{otherwise}\\
\end{array} ;\; \right.\end{equation} $$\forall_{s,t}  \;\;\; p(s,t)=\frac{1}{4};
 $$

It is a known result that $\omega_c^{\textup{CHSH}}=\frac{3}{4}$. One example of a deterministic strategy that maximizes $\textup{ W}_{\mathcal{C}}$ is given by 
\begin{equation*}
F(s,t) \equiv \{ f(s), g(t) \} \;\;\;\textup{where,} \;\;\\ \forall_{s,t}\; f(s)=g(t)=0,
    \end{equation*}
which means the players ignore the questions and always answer with $0$. In the next section we see how to construct the Tsirelson bound for the CHSH game.

It is also worth mentioning that despite this work focusing mainly on simple 2-player classical XOR games, there are very technically demanding generalizations in the literature regarding XOR games. For instance, in \cite{Ambainis}  XOR games with a large number of players are considered and their classical and quantum values are calculated, when under the restriction that the questions themselves are also bits. In \cite{Vidick} a specific class of ($n\geq 3$)-player XOR games is described where no restriction is imposed over the set of questions, which is assumed to have a cardinality $N^2$ -- for such games the authors prove that the ratio between the quantum  and classical biases\footnote{The quantum/classical bias is the difference between the quantum/classical value and the winning probability offered by a trivial random answer -- it is a standard way to measure the quantum over classical advantage in non-local games.} is  of the order $\sqrt{N}$. Another important generalization was introduced in \cite{Regev} where the notion of  quantum XOR games was proposed, in which the \textit{questions } and \textit{answers} are allowed to be quantum states.

\subsection{Best quantum strategies for 2-player XOR  games}

In Cleve et al. \cite{Cleve}, two powerful results were proven that we are going to use explicitly in the construction. These  results specify some common features that the best quantum strategies for XOR games share. The results are not explicitly stated like so in the original paper, but they are equivalent to the following:
\begin{itemize}
    \item If a non-local game is an XOR game, then the best strategy will be one where the POVMs are just projective measurements;
    \item For 2-player XOR games of sufficiently small dimensions,  the best strategy will be always realizable if Alice and Bob share an \textit{ebit} of information;
\end{itemize}{}

Based on these results we are motivated to define a generic strategy for 2-player XOR games, which abides in the most general way to the previous restrictions. As such, for any 2-player XOR game of sufficiently small dimensions, we put forward the best strategy,

\begin{equation} \label{recipe}
\begin{split}
\ket{\psi}\rightarrow\ket{B_{x y}} = & \frac{1}{\sqrt{2}}\big( \ket{0y}+(-1)^x\ket{1\Bar{y}} \big);\; x,y\in\{0,1\}; \\ \\ \hspace{-1cm}
 & \Hat{\Pi}_A \otimes \Hat{\Pi}_B \rightarrow \Hat{\mathcal{A}}\otimes \Hat{\mathcal{B}};
\end{split}
\end{equation}
With $\ket{B_{x y}}$ representing any of the four Bell states,  
\begin{equation} \label{eq1}
\begin{split}
\ket{B_{00}}= \frac{1}{\sqrt{2}}\big( \ket{00}+\ket{11} \big)=\ket{\phi^{+}},\\
\ket{B_{10}}= \frac{1}{\sqrt{2}}\big( \ket{00}-\ket{11} \big)=\ket{\phi^{-}},\\
\ket{B_{01}}= \frac{1}{\sqrt{2}}\big( \ket{01}+\ket{10} \big)=\ket{\psi^{+}},\\
\ket{B_{11}}= \frac{1}{\sqrt{2}}\big( \ket{01}-\ket{10} \big)=\ket{\psi^{-}},\\
\end{split}
\end{equation}

and, \begin{equation}
\label{projection}
\Hat{\mathcal{A}}\equiv \frac{1}{\sqrt{2}}\Hat{P}\Hat{R}(\alpha_s),\;\; \Hat{\mathcal{B}}\equiv \frac{1 }{\sqrt{2}} \Hat{P}\Hat{R}(\beta_t),
\end{equation}

 with $\Hat{P}=\ket{0}\bra{0}-\ket{1}\bra{1}$,  the projection to the computational basis and $\Hat{R}({\alpha_s})$,$\Hat{R}({\beta}_t)$ the  rotation operations that Alice and Bob apply, respectively, which arguments depend explicitly on the inputs they receive, $s$ for Alice and $t$ for Bob. The generic operator $\Hat{R}(\theta)$ acts like,
$$
\Hat{R}(\theta)\ket{y}=\cos(\theta)\ket{y}+(-1)^y\sin(\theta)\ket{\Bar{y}}, \;y \in \{0,1\}
$$

From (\ref{probq}), we get that the general expression which gives the probabilities of Alice's and Bob's answers ($a$ and $b$) is given by, 

\begin{equation*}
    \bra{\psi} (\Hat{\Pi}_A \otimes \Hat{\Pi}_B)^{\dagger}(\Hat{\Pi}_A \otimes \Hat{\Pi}_B) \ket{\psi},
\end{equation*}

which upon substitution from (\ref{recipe}) yields,
\begin{equation*}
    \bra{B_{xy}} (\hat{\mathcal{A}}\otimes \hat{\mathcal{B}})^{\dagger}(\hat{\mathcal{A}} \otimes \hat{\mathcal{B}} ) \ket{B_{xy}}.
\end{equation*} The previous expression represents in  closed form the best quantum strategy for a simple 2-player XOR game.  To understand why, let us substitute the operators $\mathcal{A}$ and $\mathcal{B}$ explicitly with (\ref{projection}), and work trough the algebra, to get to the equivalent expression,\begin{equation}
\label{expect}
    \frac{1}{{2}}\bra{B_{xy}}\big(\Hat{R}(\alpha_s)^{\dagger} \otimes  \Hat{R}(\beta_t)^{\dagger}\big)(\Hat{P} \otimes \Hat{P}) \big(\Hat{R}(\alpha_s) \otimes  \Hat{R}(\beta_t)\big)\ket{B_{xy}}
\end{equation}

Now we proceed to show a useful expression -- the most general state, after arbitrary rotations have been applied by Alice and Bob onto a shared Bell state, 
\begin{multline}
\label{rot}
  [\Hat{R}(\alpha_s) \otimes  \Hat{R}(\beta_t)]\ket{B_{xy}} =  
 \cos{\big(\alpha_s+\beta_t\cdot(-1)^{x\oplus\Bar{y}}}\big)\ket{B_{xy}} +\\(-1)^{\Bar{x}}\sin{\big(\alpha_s+\beta_t\cdot(-1)^{x\oplus\Bar{y}}}\big)\ket{B_{\Bar{x}\Bar{y}}}.
\end{multline} 

Let us also define a mapping, between the Hilbert space of dimension 4, spanned by the 4 Bell states, and a 2 dimensional Hilbert space spanned by parity base states, $\ket{a\oplus b=0}$ and $\ket{a\oplus b=1}$,  

\begin{table}[ht!]
\centering
\begin{tabular}{clc}
\multicolumn{2}{c}{$\ket{\phi^{+}}\hspace{0.3cm}\vspace{0.1cm}$} & \multirow{2}{*}{$\!\!\!\!\!\Bigg\}\longmapsto{\ket{a\oplus b=0}};$} \\
\multicolumn{2}{c}{$\ket{\phi^{-}}\hspace{0.3cm}$} &   \\\\
\multicolumn{2}{c}{$\ket{\psi^{+}} \hspace{0.3cm}\vspace{0.1cm}$} & 
\multirow{2}{*}{$\!\!\!\!\!\Bigg\}\longmapsto{\ket{a\oplus b=1}};$} \\
\multicolumn{2}{c}{$\ket{\psi^{-}}\hspace{0.3cm}$} &                   \\              \end{tabular}
\end{table}
{Which can be represented more compactly as}
\begin{equation}
\ket{B_{xy}}\longmapsto \ket{a\oplus b = y}.
    \label{table}
\end{equation}
This is obviously motivated by the fact that if either state, $\ket{\phi^{+}}$ or $\ket{\phi^{-}}$, is shared between Alice and Bob, when they both measure the same observable (e.g polarization) in the computational basis  $\{{\ket{0},\ket{1}}\}$ they  will get the same eigenvalues as a result of sharing those states; as such, if they convert the eigenvalues to bits, the parity of the outcome will be even. Likewise, if they share $\ket{\psi^{+}}$ or $\ket{\psi^{-}}$, a joint measurement in the computational basis of a given observable will always yield different eigenvalues, and hence the parity of the outcome will be odd.

We can take (\ref{rot}), and express according to  mapping (\ref{table}), what would the rotated state look like in the 2 dimensional Hilbert space spanned by \\ $\{\ket{a\oplus b=0},\ket{a\oplus b=1}\}$. We have, 
\begin{multline}
    [\Hat{R}(\alpha_s) \otimes  \Hat{R}(\beta_t)]\ket{B_{xy}} \longmapsto \cos(\theta_{s,t})\ket{a\oplus b=0} + \sin(\theta_{s,t}){\ket{a\oplus b=1}},
    \label{rotations}
\end{multline}which shows evidently that, regardless of what Bell state is shared between Alice and Bob and also which arbitrary rotations they perform, if we apply mapping (\ref{table}), the outcome will generally be a state that in the space spanned by \{$\ket{a\oplus b=0}$,$\ket{a\oplus b=1}$\}, is a superposition of the base states, $\ket{a\oplus b=0}$ and  $\ket{a\oplus b=1}$. The  amplitude coefficients of the superposition are trigonometric functions of an argument, $\theta_{s,t}$, which depends on the inputs $s$ and $t$, and also varies depending on the Bell state shared. This means that the information of which Bell state Alice and Bob share must be present in the coefficients.

The following table shows exactly what arguments are inside the functions for all four possible Bell states,

\begin{center}
\begin{table}[ht!]
\centering
\begin{tabular}{clcl}
\hline
\multicolumn{2}{c}{Bell state} & \multicolumn{2}{c}{$\theta_{s,t}$} \\ \hline
\multicolumn{2}{c}{$\ket{\phi^{+}} \vspace{0.05cm}$} & \multicolumn{2}{c}{$\beta_t-\alpha_s \vspace{0.05cm}$} \\ 
\multicolumn{2}{c}{$\ket{\phi^{-}\vspace{0.05cm}}$} & \multicolumn{2}{c}{$\beta_t+\alpha_s\vspace{0.05cm}$} \\ 
\multicolumn{2}{c}{$\ket{\psi^{+}}\vspace{0.05cm}$} & \multicolumn{2}{c}{$\frac{\pi}{2}+(\alpha_s+\beta_t)\vspace{0.05cm}$} \\ 
\multicolumn{2}{c}{$\ket{\psi^{-}}\vspace{0.05cm}$} & \multicolumn{2}{c}{$\frac{\pi}{2}-(\alpha_s-\beta_t)\vspace{0.05cm}$} \\ \hline
\end{tabular}
\label{table1}
\caption{The appropriate trigonometric arguments in (\ref{rotations}), for each possible Bell state. }
\end{table}
\end{center}

\vspace{-1cm}
Now we can also show that the joint projection can be written in terms of the Bell states,

\begin{equation}
\begin{split}
\Hat{P}\otimes\Hat{P}&=\big(\ket{0}\bra{0}-\ket{1}\bra{1}\big)\otimes\big(\ket{0}\bra{0}-\ket{1}\bra{1}\big)  \\ &=(\ket{\phi^{+}}\bra{\phi^{+}}+\ket{\phi^{-}}\bra{\phi^{-}}) -(\ket{\psi^{+}}\bra{\psi^{+}}+\ket{\psi^{-}}\bra{\psi^{-}})\\ &= \sum_{x,y} (-1)^y\ket{B_{xy}}\bra{B_{xy}}
\end{split}{}
    \end{equation}
and according to (\ref{table}) it follows that in the subspace spanned by the parity base states the projection operator is mapped to, 
\begin{multline}
\Hat{P}\otimes\Hat{P} \longmapsto \sum_y  (-1)^y \; 2\ket{a\oplus b=y}\bra{a\oplus b =y}  \\ =2\big(\ket{a\oplus b=0}\bra{a\oplus b =0}-\ket{a\oplus b=1}\bra{a\oplus b=1}\big).
\label{operatormapping}
\end{multline}

Taking (\ref{rotations}) and its conjugate, 
along with the operator mapping  (\ref{operatormapping}), we are now able to see that expression (\ref{expect}), which is an expression in the 4-dimensional Hilbert space, 
is mapped to the following expression in the 2-dimensional Hilbert space spanned by the parity base states,

\begin{equation}
\label{paritymeasure}
   \Big(\bra{E}\cos(\theta_{s,t}) + {\bra{O}}\sin(\theta_{s,t})\Big)\Big(\ket{E}\bra{E}-\ket{O}\bra{O}\Big)\Big(\cos(\theta_{s,t})\ket{E} + \sin(\theta_{s,t}){\ket{O}}\Big),
\end{equation}

where $\ket{E}$ is a short notation for the even parity state, $\ket{a\oplus b=0}$, and likewise $\ket{O}$ for the odd state, $\ket{a\oplus b=1}$.

Expression (\ref{paritymeasure}) represents a measurement of an observable which eigenvalues determine the parity of the individual measurement outcomes that Alice and Bob get. Since the state is in a superposition of the even and odd base states, it means that when Alice and Bob jointly measure the state they will get, 
\begin{gather*}
    \ket{a\oplus b =0}\; \textup{with probability}\; \cos^{2}{(\theta_{s,t})}; \\ 
\ket{a\oplus b=1}\; \textup{with probability}\;\sin^{2}{(\theta_{s,t})};
\end{gather*}
    
Operationally speaking then, the strategy boils down to Alice and Bob choosing angles $\alpha$ and $\beta$, for every possible input they can receive $s$ and $t$,  such that the argument $\theta_{s,t}$ maximizes the probability to measure the Bell state in the most convenient base state, of either even or odd parity, depending on the specific input. It is clear why this approach is the ideal strategy for XOR games, since in these kind of games the individual outputs don't matter, only their parity does. We can represent symbolically the expression that gives the probability to win any XOR game, according to this generic recipe as

\begin{equation}
\label{XORgame}
\begin{split}
   \textup{ W}_ {\mathcal{Q}}(\textup{XOR}) =\sum_{s,t} p(s,t) [\cos^2{(\theta_{s,t})}\; V^{\textup{e}}_{\textup{XOR}}(s,t)+
    \sin^2{(\theta_{s,t})} \;V^{\textup{o}}_{\textup{XOR}}(s,t)], \end{split}{} \end{equation} where,
    
    \begin{equation}
        \begin{split}
            V^{\textup{e}}_{\textup{XOR}}(s,t)  =\left\{
\begin{array}{ll}
      1, \hspace{.1cm}\textup{if }\hspace{.1cm}f(s,t)=0,\\
      0, \hspace{.1cm}\textup{otherwise}\\
\end{array} 
\right.    \\ \\
             V^{\textup{o}}_{\textup{XOR}}(s,t) = \left\{
\begin{array}{ll}
      1, \hspace{.1cm}\textup{if }\hspace{.1cm}f(s,t)=1, \\
      0, \hspace{.1cm}\textup{otherwise}\\
\end{array} 
\right.    
        \end{split}{}
    \end{equation}{}
   
such that,
\begin{equation}
\label{MAX}
    \omega^{\textup{XOR}}_q\equiv \textup{Max}\{\textup{W}_{\mathcal{Q}}(\textup{XOR})\},
\end{equation}{}

which means that finding the {quantum value} for a 2-player XOR game can be reduced to solving the maximum value problem (\ref{MAX}).

\section{Simple 2-player XOR games}

The word \textit{simple} has been used throughout the paper, but exactly in what way are these 2-player XOR games {simple}? What is meant by \textit{simple} is that the number of  possible game configurations is small enough such that we can either analytically or numerically solve the maximum value problem (\ref{MAX}), for some XOR game with a general winning probability given by expression (\ref{XORgame}). Since the answers in XOR games are necessarily bits, this means that this restriction on configurations is translated to a restriction on the set of questions, in other words, a simple 2-player XOR game is a 2-player XOR game for which the cardinality of the set of questions allowed is not so big as to render the solution of (\ref{MAX}) impossible. Now we will compute the quantum value for some examples of such simple 2-player XOR games by employing the previously showed construction, i.e. using (\ref{XORgame}) to write the winning probability for said games, and solving their respective maximum value problems (\ref{MAX}).

\subsubsection*{Quantum Value for the CHSH game} We write the predicate of the game (\ref{predicateCHSH}) once again,
\begin{equation}
V(ab| st)_{\textup{CHSH}} =\left\{
\begin{array}{ll}
      1, \hspace{.1cm}\textup{if }\hspace{.1cm}s\cdot t=a\oplus b\\
      0, \hspace{.1cm}\textup{otherwise}\\
\end{array} ;\; \right.\end{equation} $$\forall_{s,t}  \;\;\; p(s,t)=\frac{1}{4},
 $$

and now we shall write the probability of winning the game explicitly for all possible game configurations, using expression (\ref{XORgame}),

\begin{multline}
\label{maxchsh}
 \textup{W}_{\mathcal{Q}}(\textup{CHSH}) =
\frac{1}{4}\cos^{2}(\theta_{0,0})+\frac{1}{4}\cos^{2}(\theta_{0,1})+\frac{1}{4}\cos^{2}(\theta_{1,0})+\frac{1}{4}\sin^{2}(\theta_{1,1}).
\end{multline}{}
The expression states that Alice and Bob need even outcomes for the first 3 terms, which correspond to questions $(0,0)$,  $(0,1)$,  $(1,0)$, respectively, and they need odd outcomes for the last term which corresponds to question $(1,1)$. Then we should find a $\theta_{s,t}$ that maximizes (\ref{maxchsh}).

The first thing we need to do is to commit to an actual Bell state. Say, without loss of generality, that Alice and Bob share the state $\ket{\phi^{-}}$, which means according to Table 1 that, 

$$\theta_{s,t}=\alpha_s+\beta_t,$$ thus we have,

\begin{equation}
\begin{split}
    W^{\textup{CHSH}}_{\mathcal{Q}}(\alpha_0,\beta_0,\alpha_1,\beta_1)  = &\frac{1}{4}\cos^{2}(\alpha_0+\beta_0)+\frac{1}{4}\cos^{2}(\alpha_0+\beta_1)+ \\ &\frac{1}{4}\cos^{2}(\alpha_1+\beta_0)+\frac{1}{4}\sin^{2}(\alpha_1+\beta_1).
    \end{split}{}
\end{equation}

If we solve (\ref{MAX}) for this case, for instance numerically in Mathematica, we get that,

\begin{equation}
\label{CHSH value}
\textup{Max} \{ W^{\textup{CHSH}}_{\mathcal{Q}}(\alpha_0,\beta_0,\alpha_1,\beta_1) \} \approx 0.853553\ldots = \cos^2{\frac{\pi}{8}} \equiv \omega_{q}^{CHSH}\end{equation}{}\

This value is achieved by Alice and Bob, when they choose the following functions over the inputs they receive,  
\begin{equation}
\begin{split}
    \alpha_s \equiv \alpha(s) =  \frac{4\pi}{16}s-\frac{\pi}{16}\\
    \beta_t \equiv \beta(t) =\frac{4\pi}{16}t-\frac{\pi}{16},
    \end{split}{}
\end{equation}{} obviously that these functions would be different if for instance they had shared another Bell state.

\subsubsection*{Quantum Value for the Odd Cycle (OC) game} Another game used as example in the Cleve et. al paper was the Odd Cycle game, in which the players' objective is to try and convince the Referee that an odd $n$-cycle graph, $C_n$ ($n>2$) is 2-colorable i.e vertices belonging to the same edge should have different colors, which obviously can't be the case since the graph has an odd number of vertices. The game proceeds as follows -- the Referee will ask Alice and Bob, $s$ and $t$, respectively, which correspond to the vertices of the graph, from $1$ to $n$,  which color he would like to know, and Alice and Bob will answer back $a$ and $b$, which correspond to the colors appropriately chosen, according to some strategy. The answers will be obviously bits, which correspond to the coding of any two distinct colors they so choose e.g. $0= \textup{black}$ and $1= \textup{white}$. There exists another particularity in this game, which is that the questions are not entirely arbitrary, i.e. the Referee can't ask any two given vertices of the graph to the players; the questions must obey the following rule:

\begin{itemize}
\item {The vertices asked are either the same, or they share an edge and the vertex asked to Bob is clockwise after Alice's; \footnote{It is this exact rule that makes this a \textit{simple} 2-player XOR game. If there would be no restrictions on the questions, then (\ref{MAX}) would be ever harder to solve for increasing values of $n$.}}
\end{itemize}

Since Alice and Bob want to convince the Referee that the  odd n-cycle graph is 2-colorable, they will have to answer with the same color if the vertices asked are the same, and with different colors if they are different. Thus the winning condition is formalized in the following predicate,

\hspace{-0.5cm}
\begin{center}
\begin{equation}
\label{OC predicate}
\begin{split}
V(ab|& st)_{\textup{OC}}=\left\{
\begin{array}{ll}
      1, \hspace{.1cm}\textup{if }\hspace{.1cm}[s\oplus 1 = t \;(mod \; n)]=a\oplus b\\
      0, \hspace{.1cm}\textup{otherwise}\\
\end{array} ;
\right.\\  \\ \hspace{-1cm}&p(s,s=t)=\frac{1}{2};\;\; p(s,s\oplus1 =t)=\frac{1}{2};
\end{split}{}
\end{equation}{}
\end{center}{}

$[s\oplus 1= t \;(mod \; n)]$ is the truth value of the proposition
$s\oplus 1= t \;(mod \; n)$. If the proposition is false it evaluates to $0$ and it means that $s=t$ is true, which is the only other option according to the rules of the game, on the other hand if indeed  $s\oplus1=t$  is true then it evaluates to 1.

Assuming that $s\oplus 1= t$ is false, which means that $s=t$ is true, the Referee asks the same vertices to both Alice and Bob, so in order for them to win they must output the same color, which is precisely to what the condition in the predicate reduces to, $a\oplus b =0$. If $s\oplus 1= t$ is true, then the Referee is asking vertices that share an edge, so Alice and Bob must output different colors, i.e $a\oplus b =1$.

The best classical strategy that Alice and Bob can conceive is actually to agree upon a possible color configuration that maximizes their winning probability, by choosing just two vertices with a common edge to be the same color, and then stick to it. Obviously they will fail if the Referee asks for the color of such two vertices, but in general that will only happen $\frac{1}{2n}$ of the times for a $C_n$ graph, which means that

\begin{equation}
 \omega_c^{\textup{n-Odd Cycle}} = 1 -\frac{1}{2n}.
 \end{equation}
 
For instance, in the special case of a 3-cycle graph, 
\begin{equation}
 \omega_c^{\textup{3-Odd Cycle}} = \frac{5}{6}. 
\end{equation}

Figure 2 shows a specific example of a possible coloring scheme that Alice and Bob could agree upon, in the case for a $C_3$ graph, that reaches the classical value.

\begin{center}
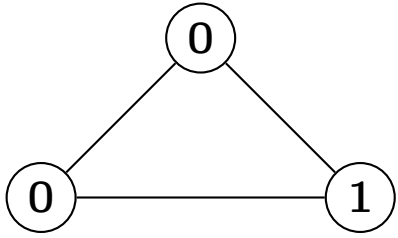
\begin{figure}[h!]
\centering
\begin{tikzpicture}[-,auto,node distance=3cm,
                    thick,main node/.style={circle,draw,font=\sffamily\Large\bfseries}]

  \node[main node] (1) {0};
  \node[main node] (3) [below left of=1] {0};
  \node[main node] (2) [below right of=1] {1};

  \path[every node/.style={font=\sffamily\small}]
    (1) edge node [left] {} (2)
       (1) edge node [right] {} (3)
       (3) edge node [right] {} (2)
      ;
\end{tikzpicture}
  \caption{If bit 0 corresponds to ``color black", and bit 1 to ``color white", for instance, the image shows a possible 2-color scheme  Alice and Bob may agree upon that maximizes their classical winning probability. The only way they can loses the game is if the Referee asks for the two ``black" vertices that share the left edge; }
    \label{odd}
\end{figure}
\end{center}

\vspace{-1cm}

Let us see how the quantum strategy goes. Alice and Bob need to answer bits whose parity is even when $s=t$ and odd when $s\oplus1=t$. From (\ref{XORgame}) we have that the best quantum strategy is
   \begin{equation}
   \label{OC}
\textup{ W}_ {\mathcal{Q}}(\textup{OC}) =
\frac{1}{2}\cos^{2}(\theta_{s,t=s})+\frac{1}{2}\sin^{2}(\theta_{s,t=s\oplus 1}).
    \end{equation}{}

If Alice and Bob share $\ket{\phi^{+}}$ then according to Table 1, (\ref{OC}) becomes 

\begin{equation}
   \label{OC2}
    \textup{W}^{\textup{OC}}_{\mathcal{Q}}(\alpha_s,\beta_t) =
\frac{1}{2}\cos^{2}(\alpha_s - \beta_{t=s})+\frac{1}{2}\sin^{2}(\alpha_s - \beta_{t=s\oplus1}).
\end{equation}{}

Now we want to solve the maximum value problem (\ref{MAX}), for the previous expression (\ref{OC2}). We will do this analytically. First, without loss of generality, assume that  $\beta_{s\oplus 1}=\beta_s - \phi_n$, where $\phi_n$ is the angle that Bob offsets $\beta_s$  (the ideal measurement orientation in the case when they receive equal inputs). To clarify -- if $\beta_s$ is the optimal orientation in which Bob performs the measurement in the situation where him and Alice receive the same input, then
$\beta_{s\oplus1}$ is the optimal orientation in which Bob does a measurement when him and Alice receive different inputs. This last orientation $\beta_{s\oplus 1}$ will now be written in terms of the other orientation and some offset angle $\phi_n$, which we make no assumption on at this point, aside from the fact that it must be something which depends on the dimension of the game. Under such considerations the probability now becomes
\begin{equation}
\begin{split}
\textup{W}^{\textup{OC}}_{\mathcal{Q}}(\alpha_s,\beta_s,\phi_n) =
\frac{1}{2}\cos^{2}(\alpha_s - \beta_s)+\frac{1}{2}\sin^{2}(\alpha_s - \beta_{s}+\phi_n).
\end{split}{}
\end{equation}{} 

Since we want to maximize the previous expression, a straightforward approach in doing so is to relate the trigonometric arguments in the following way, 

$$(\alpha_s - \beta_s)= \frac{\pi}{2}-(\alpha_s - \beta_{s}+\phi_n)$$ or equivalently,

$$(\alpha_s - \beta_s)= \frac{\pi}{4}-\frac{\phi_n}{2}.$$

Then we have that the maximum probability is 
\begin{multline}
    \label{ocmax}
 \textup{Max} \{\textup{W}^{\textup{OC}}_{\mathcal{Q}}(\alpha_s,\beta_s,\phi_n) \} =
\frac{1}{2}\cos^{2}(\frac{\pi}{4}-\frac{\phi_n}{2})+\frac{1}{2}\sin^{2}(\frac{\pi}{4}+\frac{\phi_n}{2})\\=  \frac{1}{2}\big(1+\sin{(\phi_n)}\big) \equiv \omega_q(\phi_n).
\end{multline}{}

Expression (\ref{ocmax}) shows in closed form, the {quantum value} for 
an $n$-Odd Cycle game still explicitly dependent on the generic offset angle $\phi_n$. What should $\phi_n$  be? We know that the players can't win with certainty, because the only way to do so would be to actually have a 2 color configuration of an odd cycle graph, which we know to be impossible. Bearing this in mind the following inequality comes naturally,
$$\sin{(\phi_n)}< 1\Leftrightarrow \phi_n<\frac{\pi}{2}.$$  

Also the probability to win the game should approach $1$ for ever increasing values of $n$, which translates to

$$\sin{(\omega_{n\xrightarrow \; \infty}})=1\Leftrightarrow \omega_{n\xrightarrow \; \infty}=\frac{\pi}{2}.$$

Then, the simplest expression for $\phi_n$ is, 
\begin{equation}
\label{phi}
    \phi_n = \frac{\pi}{2}(1-\frac{1}{n}),
\end{equation}{}
which in turn means that (\ref{ocmax}) becomes 

\begin{multline}
\label{OC value}
\textup{Max} \{\textup{W}^{\textup{OC}}_{\mathcal{Q}}(\alpha_s,\beta_s,\phi_n) \} = \frac{1}{2}[1+\sin{\Big(\frac{\pi}{2}(1-\frac{1}{n})\Big)}]= \cos^2(\frac{\pi}{4n}) \equiv \omega_q^{\textup{n-Odd Cycle}}.
\end{multline}{}

This result is obtained when, $$(\alpha_s-\beta_t) = \frac{\pi}{2}(1-\frac{1}{n})(s-t)+\frac{\pi}{4n},$$
which means that if, $s=t$,$$(\alpha_s-\beta_s)=\frac{\pi}{4n},$$ and if, $s\neq t$, $$(\alpha_s-\beta_{s\oplus1})=\frac{\pi}{2}-\frac{\pi}{4n}.$$

The previous arguments appear in the trigonometric functions if Alice and Bob choose the following measurements orientations, 

\begin{equation}
\begin{split}
    \alpha_s \equiv \alpha(s) = \frac{\pi}{2}(1-\frac{1}{n})s+\frac{\pi}{4n};\\
    \beta_t \equiv \beta(t) = \frac{\pi}{2}(1-\frac{1}{n})t.
    \end{split}{}
\end{equation}{}

\subsubsection*{ Quantum Value for the Entanglement Assisted Orientation in Space (EAOS) game} The EAOS game \cite{EAOS} was originated by conjuring an hypothetical  physical scenario to demonstrate the advantage of using quantum strategies in the ``real world".  The scenario is as follows -- Alice and Bob are in the poles (e.g Alice is in the South Pole and  Bob is in the North Pole) and can't communicate,
 but they want to meet in the equator line in such a way that either they
arrive  at the same point, or they arrive at points which are apart by no more than 60º along the Earth's surface, the argument being that if aided by some magnification apparatus they could still see each other in this case. 

Now, let us assume that there are 6 possible destinations to which they can arrive to, originated by setting three equally separated possible \textit{paths} (1,2,3), 120º apart, and two \textit{ways} (0,1) to go along each path. Due to the geometry of the situation, Alice and Bob win if they choose to walk along the same way for equal paths (in which case they arrive at the same destination), or walk along opposite ways for  different paths (in which case they arrive at the neighboring destinations 60º apart); see Fig.3.

\begin{center}
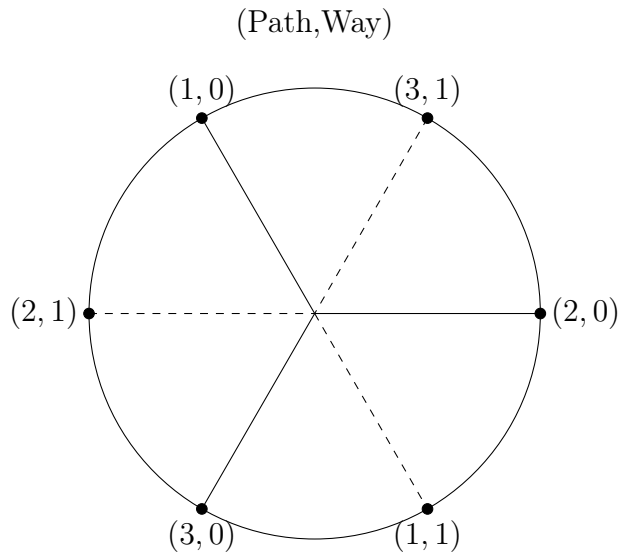
\begin{figure}[h!]
\centering
\begin{tikzpicture}
\filldraw[draw=none,fill=none] (0,3.5) circle (2pt) node[anchor=south] {(Path,Way)};
\filldraw[black] (-1.5,2.6) circle (2pt) node[anchor=south] {$(1,0)$};
\filldraw[black] (1.5,2.6) circle (2pt) node[anchor=south] {$(3,1)$};
\filldraw[black] (-1.5,-2.6) circle (2pt) node[anchor=north] {$(3,0)$};
\filldraw[black] (1.5,-2.6) circle (2pt) node[anchor=north] {$(1,1)$};
\filldraw[black] (-3,0) circle (2pt) node[anchor=east] {$(2,1)$};
\filldraw[black] (3,0) circle (2pt) node[anchor=west] {$(2,0)$};

                    \draw (0,0) circle (3cm);
                    \draw (0,0) -- (3,0) ;
                    \draw[dashed] (-3,0) -- (0,0) ;

                    \draw (-1.5,2.6) -- (0,0)  (1.5,-2.6) ;
                    \draw[dashed] (0,0)  --  (1.5,-2.6) ;
                    
                    \draw[dashed] (1.5,2.6) -- (0,0) ;
                    \draw (0,0) -- (-1.5,-2.6) ;
\end{tikzpicture}{}
  \caption{Azimuthal projection showing either the Northern (or Southern) Hemisphere, depicting at the center (where the paths cross) the North (or South) Pole. The circle corresponds to the Equator line where Alice and Bob intend on meeting, which according to the three possible paths and two ways for each path, originates into six distinct destinations represented as dots on the circle. The pair (Path, Way) specifies every possible destination.}
    \label{Fig.earth}
\end{figure}
\end{center}

\vspace{-1.5cm}
 Obviously that if both Alice and Bob had known beforehand that they would be in this scenario they could agree on a meeting point, trivializing the problem, so in order to elevate this scenario to that of a non-local game, we should assume that Alice and Bob do not have agency to pick their paths, only the ways to walk along each path, which in turn is chosen and communicated to them by a third party i.e. the Referee. Then the \textit{paths} are the questions, and the \textit{ways} are the answers of the EAOS game.
 Under such circumstances it is not terribly difficult to see that this game is an XOR game, since the winning probability depends only on the parity of the answers -- Alice and Bob win the game if they happen to  walk along the same \textit{way} for equal \textit{paths}, and opposite \textit{ways} for different \textit{paths}, regardless of what individual \textit{way} is chosen. Actually, the EAOS game bears a striking resemblance to the $n=3$ Odd Cycle game -- if instead of \textit{vertices} we have \textit{paths}, and instead of 2 possible \textit{colors} which to paint the \textit{vertices} with, we have  2 \textit{ways} to go along each \textit{path}, then it seems that, terminology aside, the set up is the same. In fact, the winning condition also seems to hold under the change of terminology -- if the \textit{paths}/\textit{vertices} are the same, then the game is won if the parity of the \textit{ways}/\textit{colors} is even. On the other hand, if the \textit{paths}/\textit{vertices} are different the \textit{ways}/\textit{colors} should have odd parity -- so we could assume that the games are equivalent, and in doing so we would be wrong. The error is in ignoring a subtle distinction in the predicates of both games. If we recall, the Odd Cycle game had an extra restriction on the way that the Referee asked the questions, 
 \begin{itemize}
    \item The vertices asked are either the same, or they share an edge and the vertex asked to Bob is clockwise after Alice's;
\end{itemize}{}
and  there is no corresponding restriction in the EAOS game. Obviously that in the EAOS game, if the paths are different, they will necessarily be adjacent to one another, but there is nothing that specifies an order between the paths each player received. If we lift this restriction from the predicate of the Odd Cycle game (\ref{OC predicate}), and write it for the special case of $n=3$, we get exactly the predicate for the EAOS game, 

  \begin{equation}
  \label{predicate EAOS1}
      V(ab| st)_{EAOS} =\left\{
\begin{array}{ll}
      1, \hspace{.1cm}\textup{if }\hspace{.1cm} [s\oplus 1= t \;(mod \; 3)]+[s\ominus 1= t \;(mod \; 3)]=a\oplus b\\
      0, \hspace{.1cm}\textup{otherwise}\\
\end{array} .
\right. 
  \end{equation}{}
So, in this scenario we evaluate the truth value of Bob's path being after $\big([s\oplus 1= t \;(mod \; 3)]\big)$ or before $\big([s\ominus 1= t \;(mod \; 3)]\big)$ Alice's, and since those are the only two possibilities when the paths are different, due to the dimensions of the game, that amounts to saying we evaluate the truth value of the paths being different, regardless of the order. Although writing the predicate in the form of (\ref{predicate EAOS1}) is useful because it illustrates the difference to the regular Odd Cycle predicate (\ref{OC predicate}), we can rewrite it in a more user friendly manner, 

\begin{equation}
  \label{predicate EAOS2}
V(ab| st)_{\textup{EAOS}} =\left\{
\begin{array}{ll}
      1, \hspace{.1cm}\textup{if }\hspace{.1cm}1-\delta_{st}=a\oplus b\\
      0, \hspace{.1cm}\textup{otherwise}\\
\end{array} ; 
\right. 
\end{equation}{}

where $\delta_{st}$ is the Kronecker delta defined as,  $$\delta_{st}=\left\{
\begin{array}{ll}
      1, \hspace{.1cm}\textup{if }\hspace{.1cm}s=t\\
      0, \hspace{.1cm}\textup{if}\hspace{.1cm}s\neq t\\
\end{array} .
\right.$$

To completely define the EAOS game we assume that the probability distribution over the set of the questions is as follows, 

$$ \;\;\;\; p(s,s=t)= p(s,s\oplus1 =t)= p(s,s\ominus1 =t)=\frac{1}{3},$$ i.e. the Referee is equally likely to demand that each player walks on any given path. At this point we have $\textup{EAOS}(V,p)$ completely defined.

The best classical strategy for the EAOS game, which was shown in \cite{EAOS}, is for Alice and Bob to agree on a deterministic mapping of the \textit{ways}
they go depending on the \textit{paths} received and allow them to share the same mapping, like in the Odd Cycle game.  Say that $f(s)=g(s)$ (i.e they share the same mapping) such that $F(s,t)\equiv  \{ f(s), g(t) \}$ is given by  $F(s,t)\equiv  \{ f(s), f(t) \}$. 
The predicate now becomes

$$V(ab| st)_{\textup{EAOS}} =\left\{
\begin{array}{ll}
      1, \hspace{.1cm}\textup{if }\hspace{.1cm}1-\delta_{st}=f(s)\oplus f(t)\\
      0, \hspace{.1cm}\textup{otherwise}\\
\end{array} . 
\right. $$

Then a possible mapping that gives the \textit{classical value} of the EAOS game is given by
$$f(1)=0; \;\;f(2)=1; \;\;f(3)=1 ;$$

Table 2 shows the winning condition evaluated for every possible combination of the outputs
that the deterministic strategy offers to Alice and Bob. The impossible conditions are in gray. It is easy to see that this strategy wins the game with a probability $$ \frac{7}{9}\equiv \omega_c(EAOS).$$

\begin{table}[h!]
\centering
\begin{tabular}{|l|l|l|l|}
\hline
   \diagbox{Alice}{Bob}    & $f(1)=0$ & $f(2)=1$ & $f(3)=1$ \\ \hline
$f(1)=0$ & $0=0$    & $1=1$    & $1=1$    \\ \hline
$f(2)=1$ & $1=1$    & $0=0$    & \cellcolor{gray!25} $1=0$    \\ \hline
$f(3)=1$ & $1=1$    & \cellcolor{gray!25} $1=0$    & $0=0$    \\ \hline
\end{tabular}
\label{table2}
\caption{ Explicit evaluations of the predicate for the shared mapping, $f(1)=0; \;\;f(2)=1; \;\;f(3)=1$;}
\end{table}

Now let us proceed to the quantum strategy. Alice and Bob need even parity outcomes when $s=t$ and odd parity outcomes when $s\oplus1=t$ and $s\ominus1=t$. Then according to (\ref{XORgame}), 
\begin{equation}
    \textup{ W}_ {\mathcal{Q}}(\textup{EAOS}) = \frac{1}{3}\cos^{2}(\theta_{s,t=s})+\frac{1}{3}\sin^{2}(\theta_{s,t=s\oplus1})+\frac{1}{3}\sin^{2}(\theta_{s,t=s\ominus1}),\end{equation}
and if the shared Bell state is $\ket{\phi^{+}}$,
\begin{equation}
     \textup{ W}_ {\mathcal{Q}}(\textup{EAOS}) = \frac{1}{3}\cos^{2}(\alpha_s - \beta_{t=s})+\frac{1}{3}\sin^{2}(\alpha_s - \beta_{t=s\oplus1})+\frac{1}{3}\sin^{2}(\alpha_s - \beta_{t=s\ominus1}). 
\end{equation}
Following the same line of reasoning as in the Odd Cycle game, we rewrite the expression such that Bob's orientations in the odd parity terms ($\beta_{t=s\oplus1},\beta_{t=s\ominus1}$), are given as functions of an ``offset angle", $\phi_3$, from his orientation in the even parity case ($\beta_{t=s}$). Due to the symmetry of the situation, we assume that the way that Bob offsets his ideal measurement orientation, in the case where the Alice's path is after Bob's, i.e $\beta_{t=s\ominus 1}$, will be the negative of the case when Alice's path is before Bob's, $\beta_{t=s\oplus 1}$. Thus we have
\begin{multline}
 \textup{ W}_ {\mathcal{Q}}^{\textup{EAOS}}(\alpha_s,\beta_s) = 
\frac{1}{3}\cos^{2}(\alpha_s - \beta_{s})+\frac{1}{3}\sin^{2}(\alpha_s - \beta_{s}+\phi_3)+ \frac{1}{3}\sin^{2}(\alpha_s - \beta_{s}-\phi_3),
\end{multline}{} where $\phi_3$ is computed from (\ref{phi}) and we get 
$$\phi_3 = \frac{\pi}{2}(1-\frac{1}{3})=\frac{\pi}{3}.$$
This in turn gives
\begin{multline}
       \textup{ W}_{\mathcal{Q}}^{\textup{EAOS}}(\alpha_s,\beta_s)  =
\frac{1}{3}\cos^{2}(\alpha_s - \beta_{s})+\frac{1}{3}\sin^{2}(\alpha_s - \beta_{s}+\frac{\pi}{3})+\frac{1}{3}\sin^{2}(\alpha_s - \beta_{s}-\frac{\pi}{3}).
\end{multline}{}
If we define $\alpha_s - \beta_{s}\equiv x$ we have the probability given as a function of $x$,

\begin{equation}
\label{EAOSP}
\textup{ W}_ {\mathcal{Q}}^{\textup{EAOS}}(x)=\frac{1}{3}\cos^{2}(x)+\frac{1}{3}\sin^{2}(x- \frac{\pi}{3})+  +\frac{1}{3}\sin^{2}(x+\frac{\pi}{3})
\end{equation}and we can compute the derivative and calculate the global maximum of the function. We get that \begin{multline}
        \textup{Max}\{  \textup{ W}_ {\mathcal{Q}}^{\textup{EAOS}}(x)\} =
\frac{1}{3}\cos^{2}(0)+\frac{1}{3}\sin^{2}(-\frac{\pi}{3})+\frac{1}{3}\sin^{2}(\frac{\pi}{3}) = \frac{5}{6} \equiv \omega_q(\textup{EAOS})
\label{EAOS value}
\end{multline}{}

This is the exact value that was achieved by the strategy presented in the original paper \cite{EAOS}, proving that the strategy is in fact the optimal quantum strategy. The strategy is achieved by setting the following measurements orientations, \begin{equation}
\begin{split}
   & \alpha_s \equiv \alpha(s) = \frac{\pi}{3}s - \frac{\pi}{3} \\&\beta_t \equiv \beta(t)=\frac{\pi}{3}t - \frac{\pi}{3}
\end{split}{}
\end{equation}{} 
\section{Conclusions}
Based on two theorems proved in \cite{Cleve}, we have seen a constructive approach to compute the Tsirelson bounds for a set of 2-player XOR games, which we called {simple} 2-player XOR games, for which finding the bound reduces to solving a maximum value problem (\ref{MAX}). The adjective \textit{simple} is a loose characterization, which means that the analytic or numerical solution of the maximum value problem that comes out of the strategy is solvable. The number of possible game configurations will, in principle, dictate the difficulty of finding the solution of (\ref{MAX}), and since in XOR games the set of answers is restricted to bits, that means a given XOR game could be \textit{simple} or not depending on the size of the set of questions. We used the strategy explicitly in the calculation of the Tsirelson bound for two well known examples of such simple 2-player XOR games --  the CHSH (\ref{CHSH value}) and the $n$-Odd Cycle (\ref{OC value}) bounds. Additionally, we also computed the Tsirelson bound for the EAOS game, which was calculated by hinging on the fact that its predicate  (\ref{predicate EAOS1}) could be retrieved by relaxing the predicate of the 3-Odd Cycle game and, as such, was also a simple 2-player XOR game where our constructive strategy was valid. Solving the maximum value problem (\ref{MAX}) for the winning probability (\ref{EAOSP}) we got the EAOS bound (\ref{EAOS value}), which was calculated for the first time since the game was introduced in \cite{EAOS}. Furthermore, we can also conclude that since the  value of the EAOS bound (\ref{EAOS value}) is achieved by the strategy presented in the original paper, there is no better quantum strategy to win the EAOS game.

\section{Acknowledgements}
The author would like to thank N. Paunković, who first introduced him to the EAOS game, and  Instituto de Telecomunicaç\~oes (IT) Research Unit, Ref. UID/EEA/50008/2019, funded by Fundação para a Ciência e Tecnologia (FCT). The author acknowledges funding from FCT through the DP-PMI  Grant PD/BD/128636/2017.

\bibliographystyle{plain}

\onecolumn\newpage



\end{document}